\documentclass[journal]{IEEEtran}
\usepackage{lineno,hyperref}

\usepackage{amsmath,amssymb,graphicx,amsthm, booktabs}
\usepackage{tikz}
\usepackage{tkz-graph}
\usepackage[noend]{algpseudocode}
\usepackage[ruled,vlined]{algorithm2e}
\usepackage{subfigure}
\usetikzlibrary{matrix}
\usetikzlibrary{shapes,arrows}
\usepackage{multirow}
\newenvironment{definition}[1][Definition]{\begin{trivlist}
\item[\hskip \labelsep {\bfseries #1}]}{\end{trivlist}}

\DeclareMathOperator*{\argmax}{\arg\!\max}
\ifCLASSINFOpdf
\else
\fi
\begin{document}
\title{A GRAPH DOWNSAMPLING TECHNIQUE BASED On GRAPH FOURIER TRANSFORM}

\author{Nileshkumar Vaishnav and Aditya Tatu\\ DAIICT, Gandhinagar, India.}
\maketitle

\begin{abstract}
In this paper, we provide a Graph Fourier Transform based approach to downsample signals on graphs. For bandlimited signals on a graph, a test is provided to identify whether signal reconstruction is possible from the given downsampled signal. Moreover, if the signal is not bandlimited, we provide a quality measure for comparing different downsampling schemes. Using this quality measure, we propose a greedy downsampling algorithm. Most of the prevailing approaches consider undirected graphs, and exploit the topological properties of the graph in order to downsample the grid, while the proposed method exploits spectral properties of graph signals, and is applicable to directed graphs, undirected graphs, and graphs with negative edge-weights. We provide several experiments demonstrating our downsampling scheme, and compare our quality measure with measures like normalized cuts.
\end{abstract}

 \ifCLASSOPTIONpeerreview
 \begin{center} \bfseries EDICS Category: 3-BBND \end{center}
 \fi
\IEEEpeerreviewmaketitle

\section{Introduction}
   There are many applications, where the domain of the measured data can be modeled as a graph. Examples of such data include weather data, seismic activity data, sensor networks data, social network data, transportation data. Given the large scope of applications\cite{Shuman2013}, analysis and processing of signals on graph is important. Signals on graph often come from a nonuniform grid, and there is no natural ordering of the vertices; rather the inter-relations between vertices is important. Defining concepts such as shift, Fourier transform and convolution is not trivial and diverges greatly from similar concepts defined for uniform signals.

    Formally, a graph is a collection of vertices with a given relation structure between the vertices. The relation between vertices is given by a matrix called the graph adjacency matrix. For an unweighted graph, the adjacency matrix has binary entries. For an undirected graph, the adjacency matrix is symmetric. Traditionally, spectral properties of graph signals are derived using \emph{graph Laplacian}. The study of eigenvalues and eigenvectors of graph Laplacian is called \textit{Spectral Graph Theory}\cite{Chung1996}. A recent approach\cite{Sandryhaila2013a} indicates that the spectral analysis of graph signals can also be carried out effectively using the graph adjacency matrix. This approach allows us to work with signals on directed graphs, which is not possible with Graph Laplacian based approach.
    
    Often we encounter signals on graph which are \emph{smooth} in nature. Such signals exhibit low-pass behavior in spectral domain. When a graph signal does not contain frequency content above a certain cut-off frequency, it is called a \emph{bandlimited} signal (a formal definition is provided in section 2). If the graph signal is bandlimited, it can be reconstructed from fewer samples in vertex-domain. The process of finding the collection of vertices which can reconstruct the original signal is given various names:\emph{graph coarsening}\cite{Liu2014,Gfeller2007}, \emph{site percolation}\cite{Grimmett1997}. Graph downsampling is a special case of graph-coarsening, where we reduce the nodes by an integer factor (e.g. downsampling by a factor of $2$ implies removing half vertices). Graph downsampling can be used for compression and as a building block for multiresolution analysis for signals on graph\cite{Hammond2011}. In this paper, the term \emph{downsampling} refers to downsampling by a factor of two, unless explicitly specified.
        
Recently, research on graph downsampling has gained momentum in the field of signal processing. Downsampling of a graph with respect to bandlimited signals draws in analogy from the classical uniform sampling and downsampling process. In 1-D uniform case, there is an ordered set of vertices, and the downsampling process amounts to selecting every alternate vertex from the set of vertices. Spectrally, the  selection of every alternate vertex results in folding of spectrum exactly by a factor of two. If the signal is bandlimited with upper-half of frequency content absent, then spectral folding does not introduce any aliasing. Thus, any signal which has no spectral content on the upper-half of the frequency spectrum can be recovered from the downsampled vertices without any error. Thus the spectral view of the signal coincides with the topological view in case of classical signal processing.\\
    Downsampling on graph, however, differs from the traditional view in both the domains (the vertex domain and the spectral domain). This is because of the fact that a graph does not provide any topology in which the vertices are ordered (except in special cases), hence \emph{selecting every alternate vertex} is not a meaningful operation. Moreover, the spectrum of a graph (i.e. eigenvalues of Laplacian/adjacency matrix) does not necessarily show symmetry, indicating that the spectral-folding phenomena is not the same as that in classical signal processing. Another challenge in downsampling on graphs is how to determine the inter-relations among the reduced set of vertices. The determination of new adjacency relation in the reduced graph is essential to obtain a multi-resolution on graph \cite{Zhang2015,Doerfler13,Nguyen2015}.\\
    In this paper, we obtain a sampling scheme which takes into account the spectral properties of the graph in order to downsample signals. The approach presented can be applied to both directed as well as undirected graphs. The approach is also applicable to the graphs with negative edge-weights \footnote{Negative edge-weights usually indicate a negative correlation between signals on two vertices, e.g. in a Social Network, two individuals can be connected by an inverse relation resulting in a negative edge-weight.}. In case the signal is not bandlimited, we provide a measure that allows to choose a scheme with minimum reconstruction error.
    
    The paper is organized as follows. Section 2 provides the background and related work in the field of graph-downsampling. In section 3 and 4, we provide our proposed downsampling method for band-limited and non band-limited (low pass) signals. A greedy algorithm to implement the proposed method is presented in Section 5, followed by several experiments to validate our claims in Section 6. We conclude the paper in Section 7, in which some future research directions are also listed.
\section{Related Work}
We begin by introducing notations and terms that are used frequently in the paper.
\subsection{Definitions and Notations}
A graph $G$ is denoted as $(\mathcal{V}, A)$, where $\mathcal{V}$ is the set of vertices $\{v_1, ..., v_N\}$ with a specified order and $A$ is the graph adjacency matrix which provides the relation structure between the set of vertices. For matrix $A$, each element $a_{i,j}$ is the weight connecting vertex $v_j$ to vertex $v_i$.\\
A \textit{graph signal} is defined as the vector $\bar{x} = [x_1, x_2, \cdots, x_N]^T$, where $x_i$'s are scalar values sampled on vertices $v_i$'s respectively. Thus a signal $\bar{x}$ can be thought of as an element in $\mathbb{C}^N$. For undirected graphs, \textit{Graph Laplacian} is defined as $L = D - A$, where $D$ is a diagonal matrix with $d_{i,i}$ being the sum of edge-weights connecting vertex $v_i$. \textit{Normalized Graph Laplacian} is defined as $L_n = D^{-1/2}LD^{-1/2}$. Given graph-Laplacian $L = V\Sigma V^T $, where $\Sigma$ is a diagonal matrix and $V$ is an orthogonal matrix, $V^T$ is the designated \textit{Graph Fourier Transform based on Laplacian}, denoted by $GFT_L$. Given normalized graph-Laplacian $L_n = V\Sigma V^T $, where $\Sigma$ is a diagonal matrix and $V$ is an orthogonal matrix, $V^T$ is the designated \textit{Graph Fourier Transform based on normalized graph Laplacian}, denoted as $GFT_N$. Following \cite{Sandryhaila2013a}, the matrix $V^{-1}$ in $A=V J V^{-1}$, which puts the given adjacency matrix $A$ into its Jordan Normal Form (JNF) $J$ is designated as the \textit{Adjacency matrix based Graph Fourier Transform}, denoted as $GFT_A$. In the case of $GFT_L$ and $GFT_N$, the ascending frequency order correspond to usual ascending order on the respective eigenvalues, while for $GFT_A$ the ascending frequency order corresponds to an ascending order on $|1-\frac{\lambda}{\lambda_{max}}|$, where $\lambda$ is the respective eigenvalue and $\lambda_{max}$ is the maximum eigenvalue, for details refer \cite{Sandryhaila2014a}. In this paper, unless specified, we assume Adjacency Matrix Based GFT. The terms \emph{nodes} and \emph{vetrices} are used interchangeably.

   Having defined a GFT, one can now define bandwidth of a signal. Consider an undirected graph, with $GFT_N$ as the designated GFT. For normalized graph Laplacian, all eigenvalues are real and non-negative and lie in interval $[0,2]$. The bandwidth of a signal, in this context, can be interpreted as a real number in the interval $[0,2]$\cite{Pesenson2008}. Another way to interpret bandwidth of a graph signal, is to count the number of eigenvalues which lie below a certain cut-off threshold. This number itself can be interpreted as bandwidth. We define the bandlimited of a graph signal as follows.
\begin{definition}
\emph{Bandlimited Signal On Graph} For a signal $\bar{x}$ on a given graph with GFT $\bar{b}$, if $\bar{b}(i)=0, \forall i\geq n_0$, then the signal is called bandlimited with bandwidth $n_0$.
\end{definition}
Due to different behaviors in vertex-domain and spectral-domain, the downsampling on graph can be looked at from both topological and spectral viewpoints.
    
    If the signal is bandlimited in spectral domain, then it can be downsampled without loss of data. One key problem in graph downsampling is determining a sample-set, i.e., the set of vertices from which a bandlimited signal can be recovered without any error. A method to determine the sample-set of an undirected graph is provided by Anis \emph{et~al.} \cite{Anis2014}, in which a greedy approach is used to add a vertex in every iteration to the sample-set, which provides the highest increase in bandwidth, until the cut-off threshold is reached. Due to the greedy nature of the algorithm, the sample-set so obtained is not necessarily \emph{optimal}\footnote{The meaning of optimality will be provided in later sections}. As an example, every alternate sample is not necessarily selected during downsampling of a standard 1-D uniform grid. It should be emphasized here that the sample-set (of a given cardinality) for a given bandwidth is not unique, and the algorithm indeed converges to one of those sample-sets. However, different sample-sets have different sensitivity to aliasing in case of signals which are not bandlimited, which indicates that even among sample-sets, the quality of signal-reconstruction differs. The objective in \cite{Anis2014} is to find the least number of samples (and corresponding sample-set) for a signal with given bandwidth. On the other hand, the purpose of our proposed approach is to provide a way to select the \emph{best possible} $N/2$ vertices for a graph with $N$ vertices. 
    
    The eigenvector corresponding to highest frequency is used to obtain a downsampling scheme in \cite{Biyikouglu2007}. Based on polarity of eigenvector values, two equivalent sets of downsampled vertices are obtained. Other approaches to downsample include those approaches which exploit the topological properties of the graph. One major class of graphs is called \emph{bipartite graphs}, which provide a natural way to downsample. An analysis of downsampling $k$-regular bipartite graphs is provided in \cite{Narang2011}. However, not all graphs exhibit bipartite structure, so to apply the downsampling to arbitrary graphs, a method proposed in \cite{Narang2O12} locally approximates the bipartite structure using a graph-colouring technique. On the other hand, Nguyen and Do\cite{Nguyen2015} rely on Maximum Spanning Tree(MST) of a graph in order to downsample\footnote{It should be noted here that every tree is a bipartite graph.}. A major limitation with topological approaches is that although the signal is assumed to be bandlimited in spectral domain, the actual process of finding the downsampling scheme does not take into account spectral properties of the graph in a direct way. Moreover, these approaches cannot be applied to downsample a directed graph, or a graph with negative edge-weights.\\
    Another issue in graph downsampling is measuring the quality of the affected partition on graph. Cut-index\cite{Nguyen2015} is a popular objective measure used to determine quality of the graph downsampling scheme, and is defined as the ratio of sum of edgeweights of edges to be deleted in order to disconnect two selected partitions, and the total edgeweights in the graph. A downsampling scheme with higher cut-index is considered to have better quality (and hence better signal reconstruction properties). One major issue with cut-index is that a single cut provides us two downsampling options (i.e. both partitions are considered equally good), selecting one of the two is an arbitrary choice.
\tikzstyle{every node}=[circle, draw,
                        inner sep=0pt, minimum width=4pt]
                        
To understand the issues with topological downsampling, consider the graph as shown in Figure 1, with all edgeweights set to $1$. Table 1 provides all possible combinations of downsampling the graph by a factor of $2$, with the corresponding proposed quality measure, which we refer to as $SDQM$ (details provided in Section 4),  and cut-index. Higher quality measure indicates lower reconstruction error. It should be noted from the table how the proposed quality measure captures the symmetries which are present in the graph. For example, selections \{1, 4, 6\} and \{1, 2, 5\} are topologically symmetric, and hence they have equal quality measure. The example under consideration also explains the limitation of cut-index quality measure. It can be seen from the table that sets \{1,2,5\} and \{3,4,6\} have identical cut-index measures. On the other hand, the proposed measure indicates that retaining \{3,4,6\} would lead to a better  reconstruction.\\\\
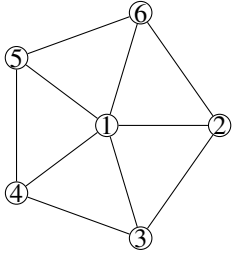
\begin{figure}
\begin{center}
\begin{tikzpicture}[scale = 0.6]
\SetGraphUnit{3} 
 %styles
\GraphInit[vstyle=Normal] 
\SetVertexNormal[Shape=circle,MinSize=0.2cm,LineWidth =1pt]
\tikzset{VertexStyle/.append} 
    \node (1) at ( 2.5, 2.5) {1}; 
    \node (2) at ( 5, 2.5) {2};
    \node (3) at ( 3.25, 0) {3};
    \node (4) at ( 0.5, 1) {4};
    \node (5) at ( 0.5, 4) {5};
    \node (6) at ( 3.25, 5) {6};

    \begin{scope}[every path/.style={-}]
       \draw (1) -- (2);
       \draw (1) -- (4); 
       \draw (1) -- (6);
       \draw (1) -- (3);
       \draw (1) -- (5);
       \draw (2) -- (3);
       \draw (3) -- (4);
       \draw (4) -- (5);
       \draw (5) -- (6);
       \draw (2) -- (6);
    \end{scope}  
\end{tikzpicture}
\end{center}
\caption{An example graph to be downsampled by a factor of 2}
\end{figure}
%\begin{table}
%\begin{tabular}{ c c c }
%\hline
%Selected Nodes & Proposed Measure & Cut-Index\\
%\hline
%$\{1,4,6\}, \{1,2,5\}, \{1,3,5\}, \{1,3,6\}, \{1,2,4\}$ & $0.1913$ & $0.7$\\  
%$\{2,3,4\}, \{4,5,6\}, \{2,3,6\}, \{2,5,6\}, \{3,4,5\}$ & $0.2539$ & $0.5$\\
%$\{1,2,3\}, \{1,2,6\}, \{1,4,5\}, \{1,3,4\}, \{1,5,6\}$ & $0.4069$ & $0.5$\\
%$\{3,5,6\}, \{2,3,5\}, \{2,4,6\}, \{3,4,6\}, \{2,4,5\}$ & $0.5257$ & $0.7$\\
%\hline
%\end{tabular}
%\caption{Proposed Quality measure for all possible Downsampling schemes for the graph shown in Figure 1}
%\end{table}
\begin{table}
\begin{tabular}{ c c c }
\hline
Set of selected nodes & $SDQM$ & Cut-Index\\
\hline
$\{1,4,6\}, \{1,2,5\}, \{1,3,5\}, \{1,3,6\}, \{1,2,4\}$ & $0.19$ & $0.7$\\  
$\{2,3,4\}, \{4,5,6\}, \{2,3,6\}, \{2,5,6\}, \{3,4,5\}$ & $0.25$ & $0.5$\\
$\{1,2,3\}, \{1,2,6\}, \{1,4,5\}, \{1,3,4\}, \{1,5,6\}$ & $0.40$ & $0.5$\\
$\{3,5,6\}, \{2,3,5\}, \{2,4,6\}, \{3,4,6\}, \{2,4,5\}$ & $0.52$ & $0.7$\\
\hline
\end{tabular}
\vspace{0.05in}
\caption{Proposed Quality measure($SDQM$) for all possible Downsampling schemes for the graph shown in Figure 1}
\end{table}
Another major issue with topological downsampling methods (e.g. MST based method) is the fact that they rely on reducing the graph to a particular structure by deleting edges, which makes them sensitive to small changes in edge-weights. For example, changing the weights of edges $(1,3), (1,4), (1,5), (2,3)$ and $(5,6)$ from $1$ to $1.01$ (i.e. a 1\% change in weights) would lead MST to drop all edges with edge-weight $1$. This in turn would result in downsampling partition $\{3,4,5\}, \{1,2,6\}$, none of which have the desired cut-index. On the other hand, the proposed downsampling method still yields \{3,5,6\} (or equivalent) and thus is less sensitive to small changes in edge-weights.

    In this paper, we emphasize on the fact that the graph signal is assumed to be bandlimited in spectral domain, hence the process of downsampling must take into account the spectral (Graph Fourier Transform) properties of the graph. With such an approach, we propose a method for downsampling that can be applied to undirected as well as directed graphs and also to the graphs with negative edge-weights. We also provide an alternative way to determine quality of the downsampling scheme, which allows us to determine which vertex-set to keep and which one to purge, i.e. both partitions may not be equivalent. The proposed algorithm optimizes this quality measure to obtain a downsampling scheme. We compare the proposed method with the spectral downsampling method (SVD based) and a topological downsampling method (MST based). %We also provide a quality-measure which allows us to compare the downsampling schemes in terms of reconstruction error. 
An analysis of the proposed approach for bipartite graphs if provided, for which the topological and proposed approach coincide; both approaches giving either of the disjoint set of vertices as the downsampled graph.
\section{Downsampling Of Bandlimited Signals And Condition For Perfect Reconstruction}
For a graph $G=(\mathcal{V},A)$ with $N$ nodes, let the GFT matrix be denoted by $F$, where $F \in \mathbb{C}^{N\times N}$. $N$ is assumed to be even as we focus on downsampling the graph by two. If a signal on this graph is bandlimited with all the energy contained in the lower half of the frequency spectrum, then the GFT of the signal is of form $[b_1, b_2, ..., b_{N/2},0, ... 0]^T$. The spectrum can be expressed as $[\bar{b}_L^T, \bar{b}_H^T]^T$, where $\bar{b}_L = [b_1, b_2, ..., b_{N/2}]$ and $\bar{b}_H = [0, ..., 0]^T$. Let $\mathcal{V}_p$ be the set of nodes to be purged and $\mathcal{V}_k$ be the set of nodes to be kept, both containing $N/2$ nodes. For a given graph signal $\bar{x}$, let $\bar{x}_k$ and $\bar{x}_p$ be the signal values taken from nodes in the sets $\mathcal{V}_k$ and $\mathcal{V}_p$ respectively. As both the sets are selections from $\mathcal{V}$, we can write, $ P_p \bar{x} = \bar{x}_p, P_k \bar{x} = \bar{x}_k$ where $P_p$ and $P_k$ are selection matrices. If we fix the order of nodes in $\mathcal{V}_k$ and in $\mathcal{V}_p$, then $P_p$ and $P_k$ are unique. We can also write,
$$ P \bar{x} = \left[\begin{array}{c} \bar{x}_k \\ \bar{x}_p \end{array}\right]$$
where $P$ is an invertible permutation matrix, with inverse $P^T$.

Similarly, we can also define selection matrices $P_L$ and $P_H$ such that $ P_L \bar{b} = \bar{b}_L, P_H \bar{b} = \bar{b}_H$.

Since $F \bar{x} = \bar{b}$, $F P^T P\bar{x} = \bar{b}$. 
$$\therefore
F_P \left[\begin{array}{c} \bar{x}_k \\ \bar{x}_p \end{array}\right] = \left[\begin{array}{c} \bar{b}_L \\ \bar{b}_H \end{array} \right]$$

where $F_P = F P^T$. If we write $F_P$ as 
\[ \left[ \begin{array}{cc}
F_1 &  F_2 \\
F_3 &  F_4 \end{array} \right],\]
then we get
\begin{equation}
\left[ \begin{array}{cc} F_1 &  F_2 \\ F_3 &  F_4 \end{array} \right] \left[\begin{array}{c} \bar{x}_k \\ \bar{x}_p \end{array}\right] = \left[\begin{array}{c} \bar{b}_L \\ \bar{b}_H \end{array} \right]
\label{eq1}
\end{equation}
where $F_1, F_2, F_3$ and $F_4$ are $\frac{N}{2} \times \frac{N}{2}$ matrices.
Given $\bar{b}_H = 0$, $\bar{x}_p$ can be uniquely determined from $\bar{x}_k$ if and only if the submatrix $F_4$ is invertible. Note that $F_4 = P_H F P_p^T$.  Moreover, using Schur Complement on the above equation, we obtain $F_{kL}$ such that
$$F_{kL}\bar{x}_k = \bar{b}_L \Leftrightarrow F_{kL} = F_1 - F_2 F_4^{-1}F_3  $$
The matrix $F_{kL}$ can be understood as the GFT on the downsampled graph. The signal on purged nodes, denoted as $\bar{x}_p$ can be recovered from $\bar{x}_k$, using the following reconstruction rule obtained from Equation \eqref{eq1}:
\begin{equation}
\bar{x}_p = -F_4^{-1}F_3 \bar{x}_k.
\label{eqn:rec}
\end{equation} 
Thus, the procedure described above, allows us to find a condition for perfect reconstruction and at the same time, provides us with the GFT on the downsampled grid. The set of samples from which the given bandlimited signal can be reproduced without any error is called a \emph{sample-set}. There can be multiple sample-sets of same cardinality for a given graph. A similar analysis for condition for perfect reconstruction of bandlimited signals is provided in \cite{Chen2015}, where the focus is solely on bandlimited signals. In this paper, we extend the same principle for non-bandlimited signals which exhibit low-pass nature.
%This gives us clues in order to form adjacency matrix for the downsampled grid. %The downsampled grids in figure() are computed using the GFT on downsampled grid ($F_{kL}$) and selected eigenvalues corresponding to lower frequencies.
\section{Downsampling Of Non-Bandlimited Lowpass Signals}
The discussion so far indicates that if the matrix $F_4 = P_H F P_p^T$ is invertible, then any bandlimited signal can be reconstructed without any error from nodes contained in $\mathcal{V}_k$. This raises a question: \emph{Are all possible node-selections with corresponding invertible $F_4$, equivalent?}
As far as bandlimited signals are concerned, all sample-sets are equivalent. However, the property of bandlimitedness is highly restrictive. In the analysis till now, we have assumed a perfectly bandlimited signal, i.e., $\|\bar{b}_H\| = 0$. However, in real-world scenarios, we often encounter situations where $0<\|\bar{b}_H\| = \epsilon << \|\bar{b}_L\|$. We refer to such signals as \emph{lowpass} signals. In this section, we will analyze this scenario which will help in obtaining an optimal downsampling scheme from the signal reconstruction point of view.
From Equation \eqref{eq1}
$$ F_3 \bar{x}_k + F_4 \bar{x}_p = \bar{b}_H
\Rightarrow \bar{x}_p = -F_4^{-1}F_3 \bar{x}_k  + F_4^{-1}\bar{b}_H.$$
The reconstruction error $e_r$ is
$$e_r = F_4^{-1}\bar{b}_H \Rightarrow \|e_r\| \leq \frac{\epsilon}{\sigma_{min}(F_4)}$$
Here, $\sigma_{min}(F_4)$ denotes the minimum singular value of $F_4$ and characterizes the sensitivity of the reconstruction error (from signal values on $\mathcal{V}_k$) to high frequency content. For a given partition $\mathcal{V}_p, \mathcal{V}_k$ the value $\sigma_{min}(F_4)$ is referred to as \emph{SVD based Downsampling Quality Measure} (abbreviated as SDQM). It should be observed here that if $SDQM = 0$, then $F_4$ is not invertible and the signal cannot be reconstructed. Maximizing SDQM reduces the upper-bound on error. As far as bandlimited signals are concerned, all downsampling schemes with $SDQM \neq 0$ are equivalent. However, when the signal is not bandlimited, they exhibit different amount of sensitivity towards the high frequency content of the signal. Thus, the goal of downsampling should be to find a sample-set that maximizes $SDQM$.

With this analysis, the problem of downsampling can be stated as the following optimization problem,
$$P_{opt} = \argmax_{P_p \in \{0,1\}^{N/2 \times N}} \{\sigma_{min}(P_H F P_p^T)\}$$
In the above optimization, $P_H$ is known (selection of high frequency components), $F$ is the GFT of graph $G$ and $P_p$ is to be found, which provides the selection of the nodes to be purged.\\
As we regard $SDQM$ as a quality measure for a given downsampling scheme, we explain the effect of this measure by an example on uniform 1-D grid (also called DFT grid\cite{Pueschel:08a}). Figure 2 shows the well-known downsampling on the grid and the resultant smaller grid for $N=6$. The optimal solutions based on $SDQM$ criteria are $\{1,3,5\}$ and $\{2,4,6\}$. Table 2 shows various selected nodes combinations and corresponding $SDQM$.
\begin{table}
\begin{center}
\begin{tabular}{ c c r }
\hline
Set of selected nodes & $SDQM$\\
\hline
$\{1,3,5\}, \{2,4,6\}$ & $0.7071$ \\  
$\{1,2,3\}, \{2,3,4\}, \{3,4,5\},...,\{6,1,2\}$ & $0.1691$\\
Rest of the combinations ($12$ in total) & $0.3568$\\
\hline
\end{tabular}
\end{center}
\caption{Quality measure for all possible downsampling schemes for graph in Figure 2. The \emph{consecutive selection} (e.g. \{1,2,3\}) shows least $SDQM$, while \emph{every alternate node selection} (e.g. \{1,3,5\}) has the largest $SDQM$.}
\end{table}
\begin{figure}
\centering
\begin{subfigure}{}
\begin{tikzpicture}[scale = 0.6]
    \node (1) at ( 0, 0) {s}; 
    \node (2) at ( 1, 0) {};
    \node (3) at ( 2, 0) {s};
    \node (4) at ( 3, 0) {};
    \node (5) at ( 4, 0) {s};
    \node (6) at ( 5, 0) {};

    \begin{scope}[every path/.style={->}]
       \draw (1) -- (2);
       \draw (2) -- (3); 
       \draw (3) -- (4);
       \draw (4) -- (5);
       \draw (5) -- (6);
       \draw (6) to [bend right] (1);
    \end{scope}  
\end{tikzpicture}
\end{subfigure} \qquad
\begin{subfigure}{}
\begin{tikzpicture}[scale = 0.6]

    \node (1) at ( 0, 0) {s}; 
    \node (3) at ( 2, 0) {s};
    \node (5) at ( 4, 0) {s};

    \begin{scope}[every path/.style={->}]
       \draw (1) -- (3);
       \draw (3) -- (5); 
       \draw (5) to [bend right] (1);
    \end{scope}  
\end{tikzpicture}
\end{subfigure}
\caption{(left) A six-node directed circulant graph, (right) Corresponding downsampled graph.}
\end{figure}
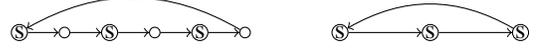

We now show that the proposed method provides expected downsampling in case of bipartite graphs.
\subsection{Analysis Of The Proposed Downsampling Method For Bipartite Graphs}
Consider a bipartite graph which has even number of nodes $N$, and equal nodes in bi-partition. An intuitive way to downsample the same is to select half of the nodes which belong to one of the partition of the bipartite structure. We use the adjacency matrix based GFT (i.e. $GFT_A$) in the given analysis. The same result can also be obtained using normalized Laplacian based GFT. The adjacency matrix of a bipartite graph is given by,
\[ A = \left[ \begin{array}{cc}
0 & B  \\
B^T & 0  \end{array} \right].\] 
where $B$ is a matrix of appropriate size. In our specific case (where we assume graph with equal number of nodes in bipartite structure), $B$ is an $\frac{N}{2} \times \frac{N}{2}$ matrix. Let the SVD (Singular Value Decomposition) of matrix $B$ be given by $ B = U \Sigma V^* $, where $U$ and $V$ are orthogonal matrices and $\Sigma$ is a diagonal matrix containing singular values of matrix in ascending order. Following \cite{Chung1996}, the matrix $A$ can be diagonalized as
$$ A = W \Lambda W^*$$
where $W = \frac{1}{\sqrt{2}}\left[ \begin{array}{cc} U & U  \\ U & -V  \end{array} \right]$ and $\Lambda = \left[ \begin{array}{cc} \Sigma & 0  \\ 0 & -\Sigma  \end{array} \right]$.
With the frequency ordering mentioned in \cite{Sandryhaila2014a}, it can be seen that the block corresponding to $-\Sigma$ contains the upper-half of frequencies. Hence, $P_H W^* = [U^* -V^*]$. As $U$ and $V$ are orthogonal matrices, an optimal selection that maximizes SDQM is selecting the first set (or the second set) of disjoint vertices.

The 1-D directed uniform graph (DFT graph) is a special case of bipartite graph and selecting every alternative node is equivalent to selecting one set of disjoint vertices of the bipartite graph.
\section{A Greedy Algorithm For Downsampling Based on SDQM}
In the formula $F_4 = P_H F P_p^T$, the matrix $P_H$ is $N/2 \times N$ rectangular matrix. Hence, $P_H F$ is also an $N/2 \times N$ rectangular matrix. Let $F_H = P_H F$, then the desired optimization turns into a column selection problem from $F_H$ such that the resultant matrix has maximum smallest singular value. A similar problem is discussed in \cite{Tropp09}, where the parameter to minimize is condition number of the selected columns from $F_H$. The number of combinations to select the columns in matrix $F_H$ are $N \choose N/2$. An exhaustive search would require computing minimum singular values $N \choose N/2$ times, which is computationally impractical for large values of $N$.

To the best of our knowledge, there is no known optimal algorithm to solve the given problem in polynomial time complexity. So, we propose a greedy strategy that may yield a suboptimal solution. The proposed greedy algorithm is summarized in \emph{Algorithm 1}. 

Let $F_4^i$ denote an $i \times N$ matrix obtained by selecting $i$ columns from $F_H$. Given $F_4^i$, $F_4^{i+1}$ is obtained by augmenting $F_4^i$ with a column from $F_H$ that maximizes the smallest singular value $F_4^{i+1}$. Iterations continue till $i=N/2$. The indices of the columns selected from $F_H$ forms the set $\mathcal{V}_p$, the set of vertices to be purged.

As discussed in section 4, the GFT for bipartite graph has two sets of equal-norm orthogonal columns in the upper-half of frequency spectrum. This allows a single orthogonal column being selected at every iteration of greedy algorithm, eventually converging to the optimal solution for any bipartite graph with equal number of nodes in each partition.
\begin{algorithm}
\SetKwInOut{Input}{Input}
\SetKwInOut{Output}{Output}
%\begin{algorithmic}
% \Input{$F_H, N$}
% \Output{$\mathcal{V}_k$}
 \caption{Downsampling On Graphs Using GFT}
% \KwData{$F_H, N$}
% \KwResult{$\mathcal{V}_k$}
 \Input{$F_H, N$}
 \Output{$\mathcal{V}_k$}
 \textbf{Procedure:} DownSample\\
 $i \gets 1$\\
 $\mathcal{V}_k \gets \{1, ..., N\}$\\
 $\mathcal{V}_p \gets \{\}$\\
 \While{$i \leq N/2$}{
  $N_d = getNodeToDelete(F_H,\mathcal{V}_k, \mathcal{V}_p)$\\
  $\mathcal{V}_p \gets \mathcal{V}_p \cup \{N_d\}$\\
  $\mathcal{V}_k \gets \mathcal{V}_k - \{N_d\}$\\
  $i \gets i+1$
 }
 return $\mathcal{V}_k$\\
% $\ $
 \line(1,0){250}\\
% $\mbox{ }$\vspace{0.1in}
% \KwData{$F_H, \mathcal{V}_k, \mathcal{V}_p$}
% \KwResult{$index$}
 \Input{$F_H, \mathcal{V}_k, \mathcal{V}_p$}
 \Output{$index$}
 \textbf{Procedure:} getNodeToDelete\\
 Array $minSVD$\\
 \ForAll{$i \in \mathcal{V}_k$}{
     $F_{iter} \gets columns\ from\ F_H\ given\ by\ \mathcal{V}_p \cup \{i\}$\\
     $minSVD(i) \gets \sigma_{min}(F_{iter})$
  }
  $index = \argmax_{i}\{minSVD\}$\\
  return $index$\\
\end{algorithm}
\section{Experimental Validation}
In this section, we apply Algorithm 1 to solve the problem of downsampling for undirected and directed graphs. The measure of quality of downsampling scheme is given by the reconstruction error, from the downsampled graph to the original graph. In subsection 6.1, we observe the effect of presence of negative edges on various downsampling schemes. In subsections 6.2 and 6.3, we downsample undirected and directed graphs respectively, and compare the reconstruction errors with existing downsampling schemes. Downsampling of DCT-graphs is used in JPEG image compression standard for the chrominance components of an image. In subsection 6.4, we demonstrate that the downsampling scheme for DCT-graphs obtained using the $SDQM$ measure outperforms the conventional downsampling.
\subsection{Downsampling Random Graphs}
In this experiment, we randomly generate graphs with $|\mathcal{V}| = 100$. We conduct the experiment for graphs with non-negative edge-weights and for graphs which have negative as well as positive edge-weights. For non-negative weights, each entry of adjacency matrix is drawn from a uniform distribution $U(0,1)$. For adjacency matrix with negative weights, each entry is drawn from Gaussian distribution $\mathcal{N}(0,1)$. The adjacency matrix thus obtained is made sparse by sparsity ratio in range of $2\%-30\%$. $1000$ instances of such matrices are generated for non-negative and negative-positive each. Table 3 summarizes average SDQM and cut-index measures for the trial using MST based approach, spectral approach and proposed approach (i.e. Algorithm 1).\\
One can observe from the table that presence of negative weights deteriorates performance of both MST based and spectral approach according to SDQM and cut-index measures. In case of spectral method, the difference is significant. This can be explained by the fact that the spectral method attempts to affect a max-cut on the given graph, hence negative edge-weights adversely affects the performance. On the other hand, the proposed approach, while optimizing SDQM also maintains cut-index comparable to MST based approach. One more remarkable feature is that the performance of proposed approach is unaffected by introduction of negative edge-weights. This experiment establishes that the proposed approach maximizes SDQM while maintaining a high cut-index and at the same time, it can also process graphs with negative edge-weights.\\ 
\begin{table}
\begin{center}
\begin{tabular}{|c|ccc|}
\hline
      & MST   &    Spectral  &      Proposed\\
\hline
\multicolumn{4}{|c|}{Nonnegative Edge-weights} \\
\hline
SDQM:& 0.0196  &   0.0178 &    \textbf{0.1449}\\
Cut Index:  & 0.5718  &   0.6158  &   \textbf{0.5721}\\
\hline
\multicolumn{4}{|c|}{Negative And Positive Edge-weights} \\
\hline
SDQM: & 0.0162   &   0.0119 &  \textbf{0.1555}\\
Cut Index: &  0.5710    &  0.5047 &  \textbf{0.6037}\\
\hline
\end{tabular}
\end{center}
\caption{SDQM and cut-index for random undirected graphs}
\end{table}
\subsection{Downsampling Undirected Graphs}
The data used in the experiment is temperature data from weather stations, publicly available on \cite{tempdata2014}. From the database, we consider $196$ nodes from which directed and undirected graphs are constructed. Data for year 2014 is considered with data available on all nodes for $365$ days. Thus, we have $365$ graph signals with number of nodes being $196$. To construct an undirected graph from the given temperature data, we use similarity measure given by statistical correlation. The diagonal entries of the correlation-matrix are all set to $0$, and the matrix is normalized with the largest eigenvalue. The matrix is then designated as the adjacency matrix of the graph. This matrix is symmetric, and hence represents an undirected graph. We diagonalize the adjacency matrix in order to obtain the GFT for the given graph\footnote{Note that the choice between graph Laplacian based GFT and adjacency matrix based GFT is arbitrary. The adjacency matrix based GFT is used here.}.\\
We obtain the downsampled grids using MST based approach, spectral method and proposed method. For each downsampled grid, we reconstruct the graph signal on purged nodes using the values on kept nodes. The reconstruction accuracy is defined as $20\log\left(\frac{\|\bar{x}\|}{\|e_r\|}\right)$. Figures 3 and 4 provide the downsampled grids (both purged and kept nodes) and reconstruction accuracy. The reconstruction accuracies indicate that the proposed algorithm outperforms both the methods.  The value of $SDQM$ for spectral downsampling approach and MST based approach are $0.02$ and $0.03$ respectively, while the same for proposed approach is $0.19$. This fact reflects directly in the reconstruction errors.\\
\begin{figure}
\begin{center}
\begin{subfigure}{}
\includegraphics[scale=0.25]{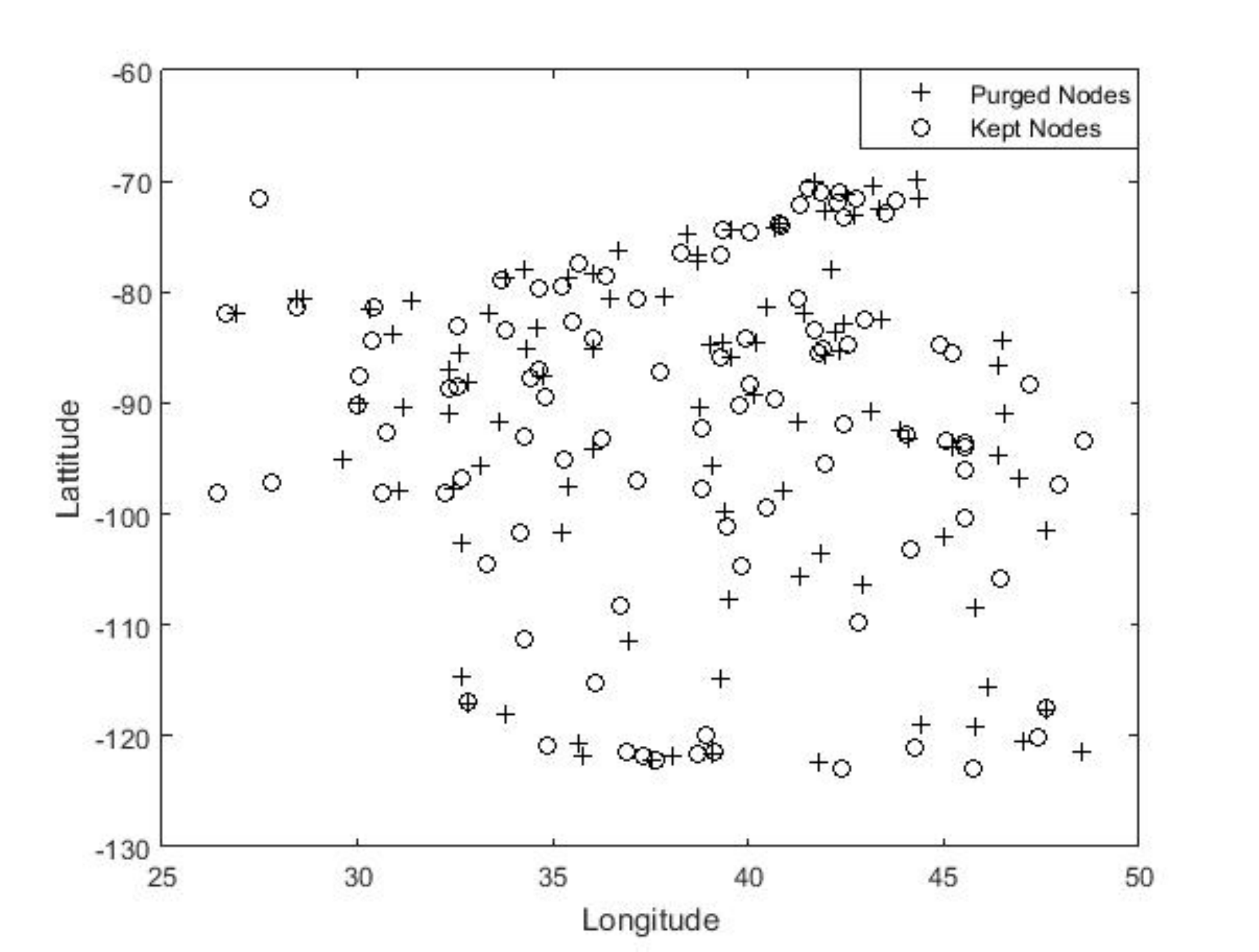}
\end{subfigure}\\
\begin{subfigure}{}
\includegraphics[scale=0.22]{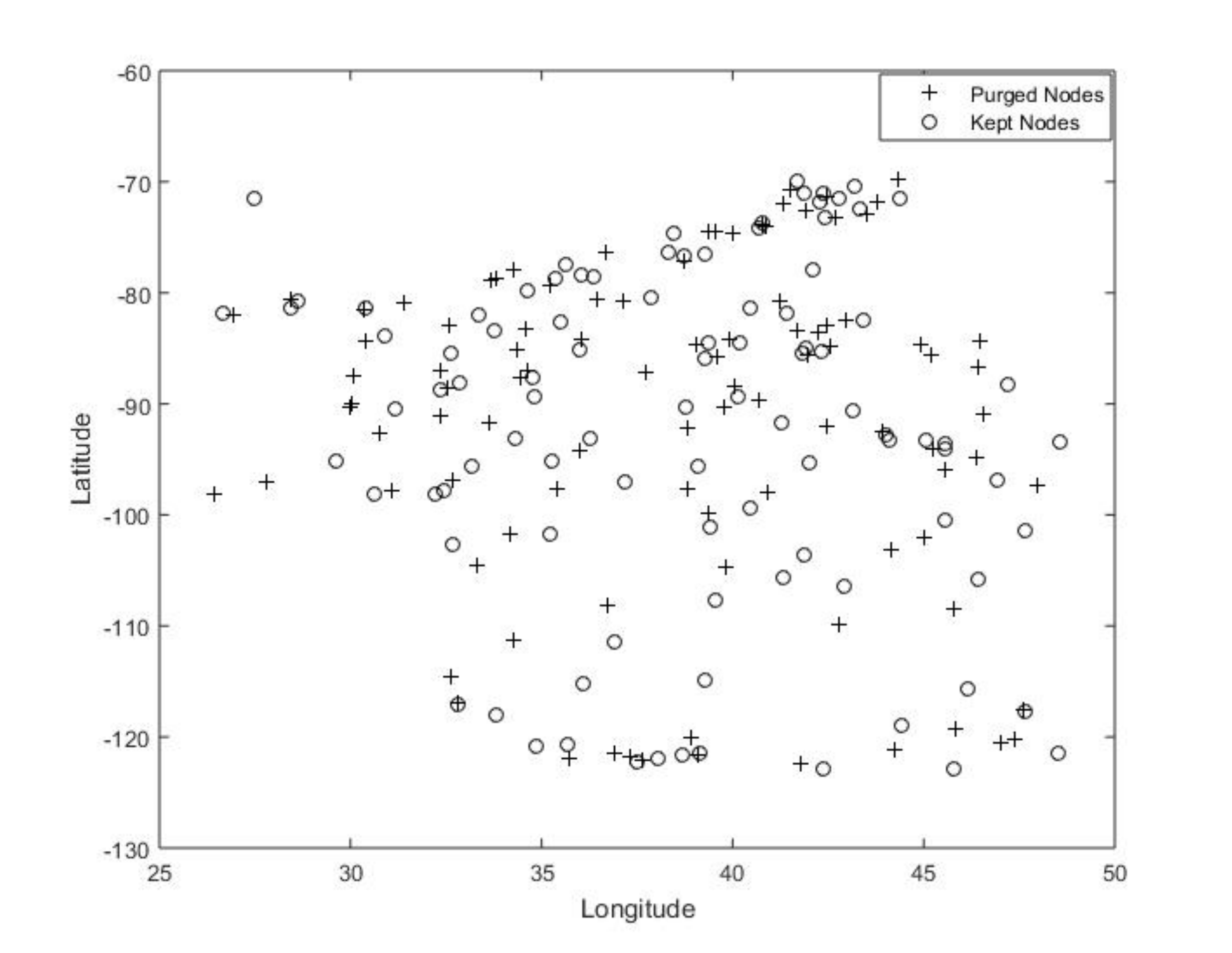}
\end{subfigure}
\caption{Result of downsampling undirected temperature data graph: (Top) MST Based approach (Bottom) SVD Based approach, $+$ denotes purged nodes, $\circ$ denotes preserved nodes.}
%\end{figure}
%\begin{figure}
\begin{subfigure}{}
\includegraphics[scale=0.2]{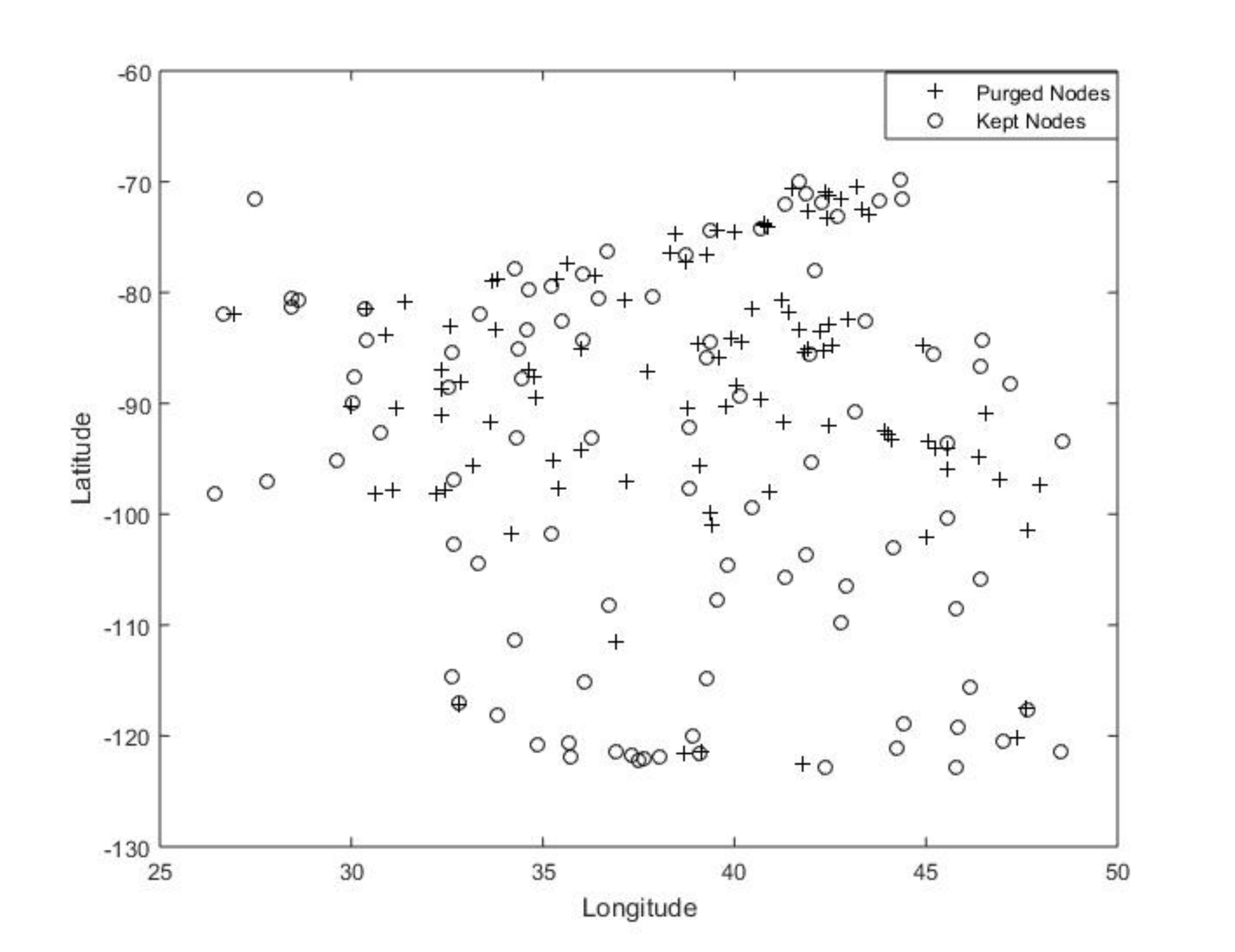}
\end{subfigure}\\
\begin{subfigure}{}
\includegraphics[scale=0.23]{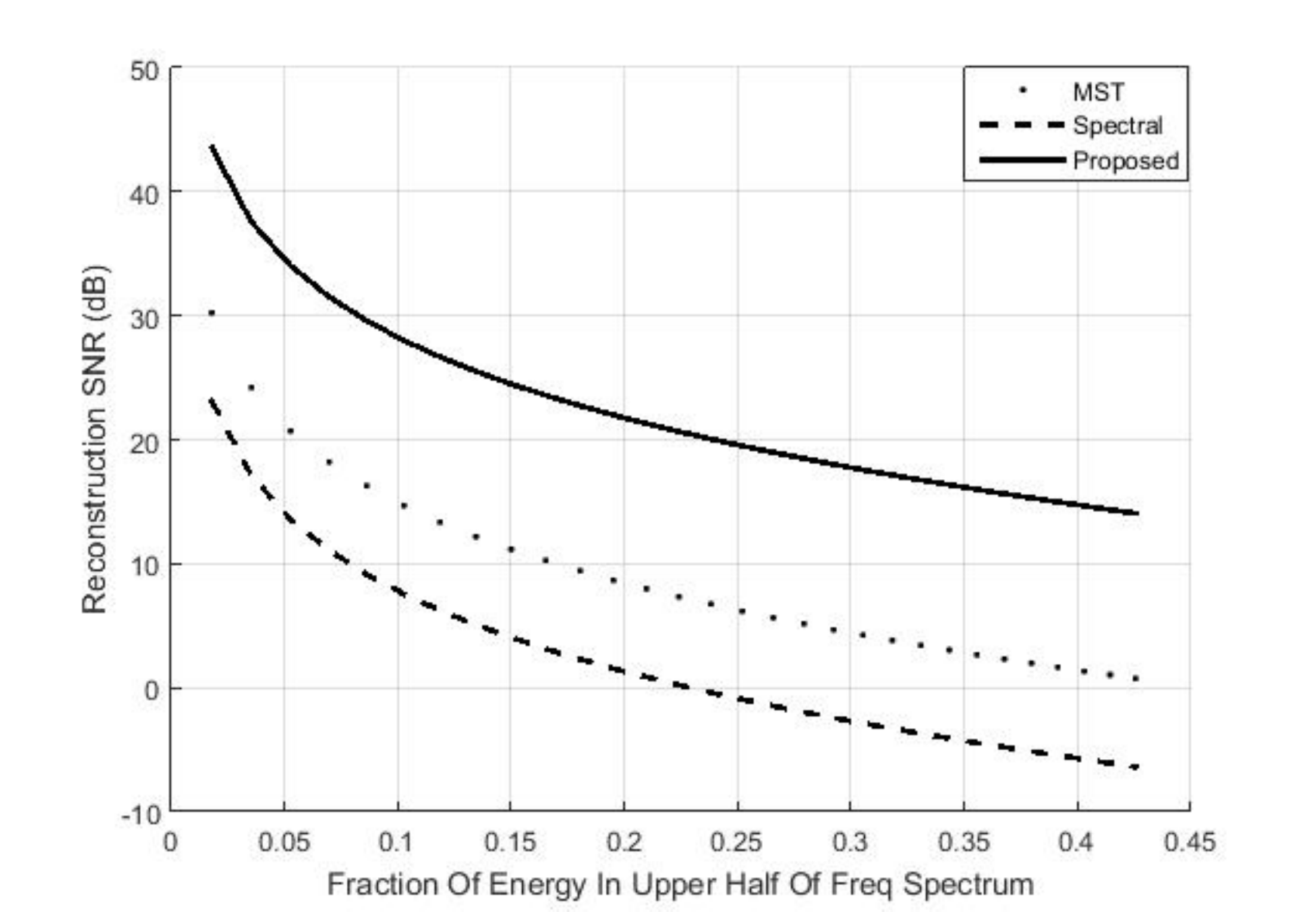}
\end{subfigure}
\caption{ (Top) Result of downsampling undirected temperature data graph using proposed method, $+$ denotes purged nodes, $\circ$ denotes preserved nodes (Bottom) Reconstruction accuracy vs High frequency content in signal}
\end{center}
\end{figure}
%\begin{figure}
%\centering
%\includegraphics[scale=0.35]{UndirectedErrorCurves}
%\caption{Reconstruction accuracy vs High frequency content in signal}
%\end{figure}
\subsection{Downsampling Directed Graphs}
For this experiment, we use the same dataset as used in Section 6.2. In order to create a directed graph for the temperature data, we first create an 8-neighborhood distance-based adjacency matrix $\tilde{A}$, whose $(i,j)$ entry is $\tilde{a}_{i,j} = e^{-\frac{dist(i,j)^2}{d_0^2}}$. Here, $d_0$ is the mean distance over entire grid. Similarly, $dist(i,j)$ is geometric (Euclidean) distance between latitude and longitude of weather stations (nodes) numbered $i$ and $j$. After this, each row of $\tilde{A}$ is normalized to have unit norm in order to obtain adjacency matrix $A$. This process makes the adjacency matrix asymmetric, hence the adjacency matrix based approach is used to obtain GFT for this graph.
 Using this GFT and Algorithm 1, we obtain a downsampling scheme on graph. For this downsampling, the reconstruction error for various levels of high-frequency content is shown in Figure 5. The value of $SDQM$ for the obtained partition is 
$0.13$.
\begin{figure}
\centering
\includegraphics[scale=0.35]{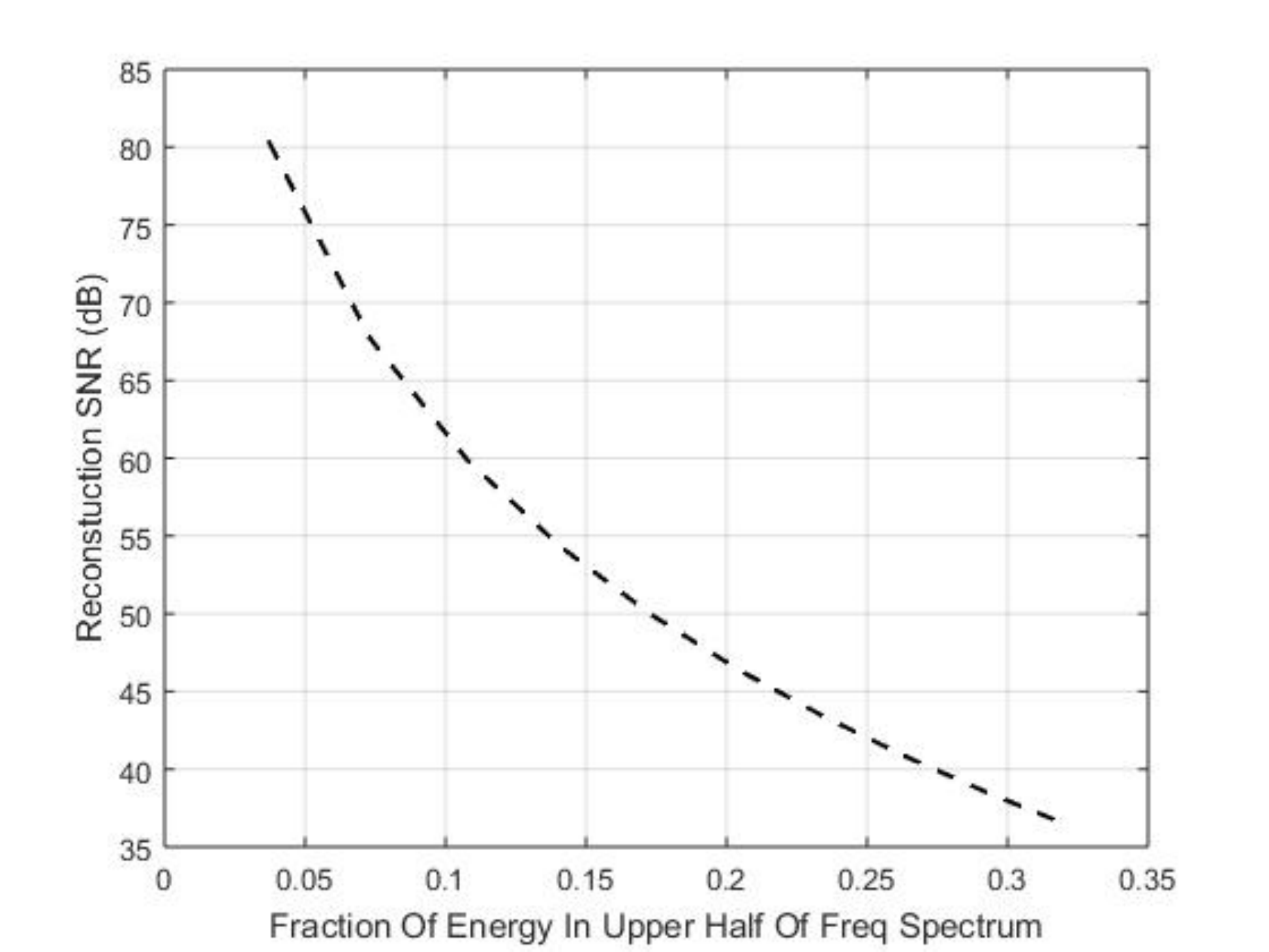}
\caption{Reconstruction accuracy vs High Frequency content in signal (Directed Graph)}
\end{figure}
%\subsection{Data Dependent Downsampling: An Example}
%In this experiment, we take a look at how the procedure proposed in the paper operates on data. We generate a synthetic signal which has first half samples with high variations and second half with low variations. We then create a 4-neighborhood value based graph for the same signal. i.e., if node $i$ and node $j$ are within 4-neighborhood of eachother, then weight connecting the two nodes is $C e^{-((s_i-s_j)^2)/\theta}$, where $\theta = 0.2$, while $s_i, s_j$ are signal values on nodes $i$ and $j$ respectively. Value of $C$ is $1$ for nearest neighbor and $0.7$ for second neighbor.\\
%If we apply the bipartite graph based downsampling approach after finding the MST, then the resultant scheme tends to pick alternate samples from the 1-D grid, in the low variation part. This means, that the downsampling would purge same number of nodes from both the high variation part and the low variation part. On the other hand, the downsampling scheme obtained using the proposed method tends to have higher number of samples in regions with high variations. In figure 4, the first half of the signal contains higher variations, and the downsampling scheme suggests that $60\%$ samples in the downsampled signal would come from first half of the signal.\\
%\begin{figure}
%\centering
%\includegraphics[scale=0.25]{datadependent}
%\caption{Variation in sampling in fast varying vs slow varyinig portions of signal}
%\end{figure}
\subsection{Downsampling DCT Graphs For Chromatic Components Of Images}
In this section, we have a look at the Discrete Cosine Transform (DCT-type II) and its graph. A detailed study on the graphs for which the DCT is GFT is provided in \cite{Puschel2008}. The graph for DCT-type II transform is given in Figure 6 (DCT-type II transform diagonalizes the adjacency matrix of this graph). The graph for DCT is undirected, and has self-loops at the end-nodes. Looking at the structure of graph, intuitively, selecting every alternate node is a good strategy for obtaining the downsampled grid. However, using the $SDQM$ measure, we find that there exists a better quality downsampling grid. For $\mathcal{V}=16$, $Algorithm \ 1$ converges to the sample-set: $\mathcal{V}_k = \{1,3,6,8,10,12,14,16\}$. Let us denote the selection of every alternate sample as set $\mathcal{V}_r = \{1, 3,5, 7, ..., 15\}$, which serves as reference for comparing the results. $SDQM$ for $\mathcal{V}_k$ is $0.4865$ while for $\mathcal{V}_r$, it is $0.4323$. According to our hypothesis, $\mathcal{V}_k$ should outperform $\mathcal{V}_r$ in signal reconstruction error. 

Downsampling of DCT-graphs is used in the JPEG compression standard for color images. In the JPEG compression standard, a color image is first converted into YUV components (Y is luminance, and U, V are chrominance components). %For efficient implementation, the image is first divided into blocks, usual size being $16\times 16$. 
As human eye is less sensitive to chrominance, every $16 \times 16$ (non-overlapping) block of U and V components, are first downsampled to $8 \times 8$ block and then 2-D DCT is applied on these blocks in order to compress the same\footnote{Note that a $16\times 16$ pixel block forms a graph that is the cartesian product of the graph given in Figure 7 with itself. Hence GFT on the $16\times 16$ node graph is the Kronecker product of GFT for graph in Figure 6, with itself. For details refer \cite{Sandryhaila2014b}}. In this experiment, we change the downsampling set from $\mathcal{V}_r$ to $\mathcal{V}_k$ and show how the sample-set $\mathcal{V}_k$ can reproduce original blocks with reduced error. For forward transform, 8-point DCT is used for $\mathcal{V}_r$ and $F_{kL}$ (see section 4) is used for $\mathcal{V}_k$. The reconstruction into $16\times 16$ blocks is performed using 16-point 2-D IDCT on the transformed blocks with appended zeros in both the cases. We select three images namely Lena, Barbara and Baboon images (all of size $192 \times 192$), which are shown in Figure 7. The blockwise average percentage errors in U and V components are provided for all three images for both the sample-sets in the table 4. The error for a single block is computed using 2-norm of the error block, and then the error is averaged over all blocks to obtain blockwise average error. It can be seen that the sample-set derived using the proposed algorithm reproduces the chromatic components with reduced error compared to standard DCT-IDCT method. The difference in SDQM explains the different performance of both schemes.\\
It should be emphasized here that the purpose of this experiment is not to provide a new method of image-compression. Rather, the purpose is to show how underlying graph structure provide non-intuitive downsampling schemes which are captured well by the proposed quality measure SDQM.
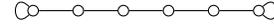
\begin{figure}
\centering
\begin{tikzpicture}[scale = 0.6]
%\SetGraphUnit{3} 
% styles
%\GraphInit[vstyle=Normal] 
%\SetVertexNormal[Shape=circle,MinSize=0.2cm,LineWidth =1pt]
%\tikzset{VertexStyle/.append} 
    \node (1) at ( 0, 0) {}; 
    \node (2) at ( 1, 0) {};
    \node (3) at ( 2, 0) {};
    \node (4) at ( 3, 0) {};
    \node (5) at ( 4, 0) {};
    \node (6) at ( 5, 0) {};

    \begin{scope}[every path/.style={-}]
       \draw (1) to [out=225,in=135,looseness=8] (1);
       \draw (1) -- (2);
       \draw (2) -- (3); 
       \draw (3) -- (4);
       \draw (4) -- (5);
       \draw (5) -- (6);
       \draw (6) to [out=45,in=315,looseness=8] (6);
    \end{scope}  
    %\path
    %(1) edge [bend right] node {} (1);
\end{tikzpicture}
\caption{Graph For DCT-type II ($|\mathcal{V}|=6$). Notice the self-loops for end-nodes.}
\end{figure}
\begin{figure}
\begin{center}
%\hspace{-0.5in}
\includegraphics[scale=0.6]{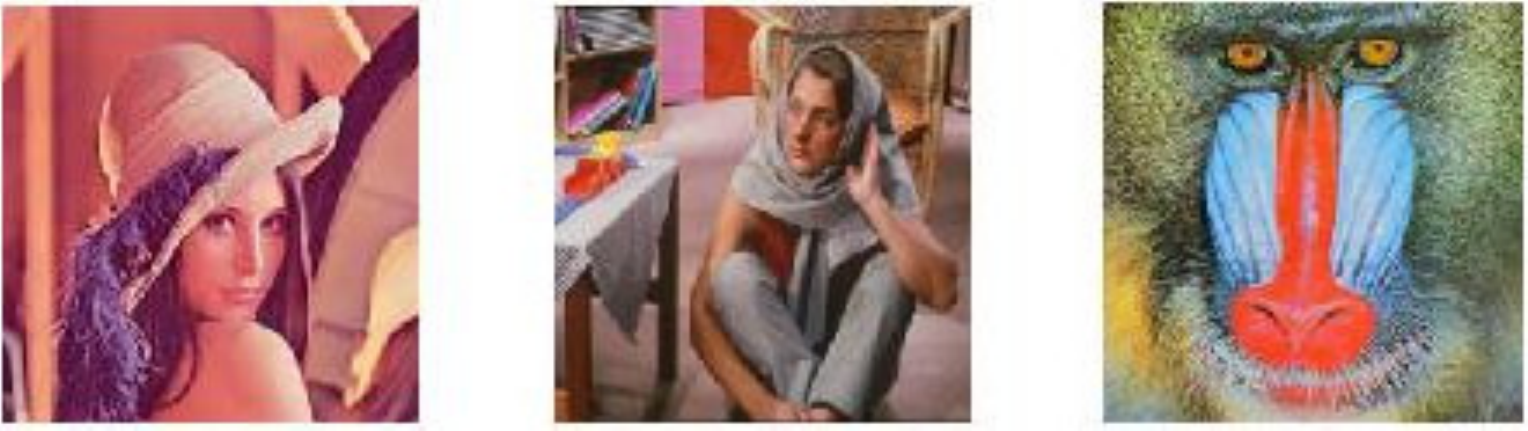}
\caption{Images used: (left) Lena (center) Barbara (right) Baboon}
\end{center}
\end{figure}  
\begin{table}[h]
\begin{center}
\begin{tabular}{|c| c c| c c| c c|}
\hline
& \multicolumn{2}{|c|}{Baboon} & \multicolumn{2}{|c|}{Barbara} & \multicolumn{2}{|c|}{Lena}\\
\hline
& $\mathcal{V}_r$ & $\mathcal{V}_k$ & $\mathcal{V}_r$ & $\mathcal{V}_k$ & $\mathcal{V}_r$ & $\mathcal{V}_k$\\
\hline
U-Component  & 4.2241 &   \textbf{3.6605}  &   1.3841  &  \textbf{0.9817} &  0.9806 &   \textbf{0.4970}\\
V-Component  & 3.6346 &   \textbf{3.0814}  &   1.2641  &  \textbf{0.8374} &  0.6843 &   \textbf{0.4108}\\
\hline
\end{tabular}
\end{center}
\caption{Blockwise Average Percentage Errors For Downsampled Images}
\end{table}
\section{Conclusion and Future Work}
To summarize, the contributions of this paper are: (1) We provide a test for finding whether a signal can be perfectly reconstructed from a given downsampled grid. (2) We propose a quality measure $SDQM$, which can be used to determine quality of a downsampling grid. (3) Based on $SDQM$, we obtain an optimization based formulation for downsampling an arbitrary graph. We also provide a greedy algorithm to solve the optimization problem. The proposed method is applicable to undirected graphs, directed graphs, and graphs with negative edge-weights.

The proposed approach is computationally challenging for large graphs. To address this issue, we are presently working on merging topological approaches with the proposed method.
We are also working towards deriving the inter-relations between the downsampled vertices based on spectral properties.
\bibliographystyle{IEEEtran}
%\bibliography{strings,refs}

\section{Comment From Authors}
This work was independently carried out by authors during the period of Nov 2015 to June 2016. Upon getting peer reviewed, it was pointed out that the work has significant overlap with work presented in \cite{Chen2015b}. The purpose of this arXiv copy is to have a reference point for nomenclature and terminology.
\bibliography{ASP_BibTex,references}
\end{document}